\documentclass[a4paper]{article}

\usepackage{stackengine}
\stackMath

\usepackage{enumitem}
\usepackage{rotating}
\usepackage{bussproofs}
\usepackage{pdflscape}
\usepackage{mathabx}
\usepackage{philex}
\usepackage{latexsym}
\usepackage{url}
\usepackage{pxfonts}
\usepackage{microtype}

\usepackage{natbib}

\newtheorem{theorem}{Theorem}

\begin{document}
\title{Two Treatments of Definite Descriptions in Intuitionist Negative Free Logic}
\author{Nils K\"urbis}
\date{}
\maketitle

\begin{center}
Published in the \emph{Bulletin of the Section of Logic} 48/4 (2019): 299–317
\url{https://doi.org/10.18778/0138-0680.48.4.04}\bigskip
\end{center}

\begin{abstract}
\noindent Sentences containing definite descriptions, expressions of the form `The $F$', can be formalised using a binary quantifier $\iota$ that forms a formula out of two predicates, where $\iota x[F, G]$ is read as `The $F$ is $G$'. This is an innovation over the usual formalisation of definite descriptions with a term forming operator. The present paper compares the two approaches. After a brief overview of the system $\mathbf{INF}^\iota$ of intuitionist negative free logic extended by such a quantifier, which was presented in \citep{kurbisiotaI}, $\mathbf{INF}^\iota$ is first compared to a system of Tennant's and an axiomatic treatment of a term forming $\iota$ operator within intuitionist negative free logic. Both systems are shown to be equivalent to the subsystem of $\mathbf{INF}^\iota$ in which the $G$ of $\iota x[F, G]$ is restricted to identity. $\mathbf{INF}^\iota$ is then compared to an intuitionist version of a system of Lambert's which in addition to the term forming operator has an operator for predicate abstraction for indicating scope distinctions. The two systems will be shown to be equivalent through a translation between their respective languages. Advantages of the present approach over the alternatives are indicated in the discussion. 
\end{abstract}


\section{Introduction}
Sentences of the form `The $F$ is $G$' can be formalised by using a binary quantifier $\iota$ that forms a formula out of two predicates as $\iota x[F, G]$. This provides an alternative to the usual way of formalising definite descriptions by means of an operator $\iota$ that forms a term out of a predicate, where $\iota x F$ is read as `The $F$'. This paper is a comparison of the two approaches. The use of the same symbol $\iota$ for the binary quantifier and the term-forming operator should not lead to confusion, as context will make clear which one is meant. In \citep{kurbisiotaI}, I presented the system $\mathbf{INF}^\iota$ of natural deduction for intuitionist negative free logic extended by the binary quantifier $\iota$ and proved a normalisation theorem for it.\footnote{For the proof-theory of term forming $\iota$ operators in the context of sequent calculi for classical logic, see \cite{andrzejfregean} and \cite{andrzejmodaldescription}.} The present paper begins with a brief overview of $\mathbf{INF}^\iota$, so that it can be read independently of the previous one. I will then compare $\mathbf{INF}^\iota$ to a system of Tennant's sketched in \citep{tennantabstraction} and \citep{tennantnatural}. Tennant provides rules of natural deduction for a term-forming $\iota$ operator within the version of intuitionist negative free logic used here. After some clarifications related to scope distinctions, it will be shown that Tennant's system is equivalent to the subsystem of $\mathbf{INF}^\iota$ in which the $G$ of $\iota x[F, G]$ is restricted to identity. Both systems are also shown to be equivalent to an axiomatic treatment of a term forming $\iota$ operator within intuitionist negative free logic. I then compare $\mathbf{INF}^\iota$ to an intuitionist version of a system proposed by Lambert in \citep{lambertfreedef}, which in addition to the term forming operator has an operator for predicate abstraction for indicating scope distinctions. Both systems are shown to be equivalent by means of a translation between their respective languages. As we go along proving these equivalences, the present paper will also illustrate the workings of the rules for the binary quantifier $\iota$ with numerous examples of deductions in $\mathbf{INF}^\iota$, and advantages of the present approach over the usual one will become apparent. In particular, in the formalisation of definite descriptions it is desirable to have a device for scope distinctions. The sole purpose of the abstraction operator in Lambert's system is as an indicator of scope. The formalism of the present system, by contrast, incorporates scope distinctions directly. Thus the formal treatment of definite descriptions with a binary quantifier is in this sense more economical than the approach using a term forming operator.\footnote{I would like to thank a referee for the \emph{Bulletin} for the careful and helpful comments.}

\section{$\mathbf{INF}^\iota$} 
Let's begin with a review of intuitionist negative free logic $\mathbf{INF}$. The rules for the propositional connectives are just those of intuitionist logic: 

\begin{center}
\AxiomC{$A$} 
\AxiomC{$B$}
\LeftLabel{$\land  I$: \ }
\BinaryInfC{$A\land  B$}
\DisplayProof\qquad\qquad 
\AxiomC{$A\land  B$}
\LeftLabel{$\land  E$: \ }
\UnaryInfC{$A$}
\DisplayProof\qquad 
\AxiomC{$A\land  B$} 
\UnaryInfC{$B$}
\DisplayProof
\end{center}

\begin{center}
\RightLabel{$_i$}
\AxiomC{ }
\UnaryInfC{$A$}
\noLine
\UnaryInfC{$\Pi$}
\noLine
\UnaryInfC{$B$}
\RightLabel{$_i$}
\LeftLabel{$\rightarrow I$: \ }
\UnaryInfC{$A\rightarrow B$}
\DisplayProof \qquad \qquad
\AxiomC{$A\rightarrow B$}
\AxiomC{$A$}
\LeftLabel{$\rightarrow E$: \ }
\BinaryInfC{$B$}
\DisplayProof
\end{center}

\begin{center}
\AxiomC{$A$}
\LeftLabel{$\lor I$: \ }
\UnaryInfC{$A\lor B$}
\DisplayProof\qquad
\AxiomC{$B$}
\UnaryInfC{$A\lor B$}
\DisplayProof\qquad\qquad
\AxiomC{$A\lor B$}
\AxiomC{}
\RightLabel{$_i$}
\UnaryInfC{$A$}
\noLine
\UnaryInfC{$\Pi$}
\noLine
\UnaryInfC{$C$}
\AxiomC{}
\RightLabel{$_i$}
\UnaryInfC{$B$}
\noLine
\UnaryInfC{$\Sigma$}
\noLine
\UnaryInfC{$C$}
\RightLabel{$_i$}
\LeftLabel{$\lor E$: \ }
\TrinaryInfC{$C$}
\DisplayProof
\end{center}

\begin{prooftree}
\AxiomC{$\bot$}
\LeftLabel{$\bot E$: \ }
\UnaryInfC{$B$}
\end{prooftree}

\noindent where the conclusion of $\bot E$ is restricted to atomic formulas. 

The rules for the quantifiers are relativised to an existence predicate: 

\begin{center}
\AxiomC{}
\RightLabel{$_i$}
\UnaryInfC{$\exists !y$}
\noLine
\UnaryInfC{$\Pi$}
\noLine
\UnaryInfC{$A_y^x$}
\LeftLabel{$\forall I:$ \ }
\RightLabel{$_i$}
\UnaryInfC{$\forall x A$}
\DisplayProof\qquad\qquad
\AxiomC{$\forall xA$}
\AxiomC{$\exists !t$}
\LeftLabel{$\forall E:$ \ }
\BinaryInfC{$A_t^x$}
\DisplayProof
\end{center} 

\noindent where in $\forall I$, $y$ is not free in any undischarged assumption of $\Pi$ except $\exists !y$, and either $y$ is the same as $x$ or $y$ is not free in $A$; and in $\forall E$, $t$ is free for $x$ in $A$. 

\begin{center}
\AxiomC{$A_t^x$}
\AxiomC{$\exists !t$}
\LeftLabel{$\exists I:$ \ }
\BinaryInfC{$\exists x A$}
\DisplayProof\qquad\qquad
\AxiomC{$\exists xA$}
\AxiomC{$\underbrace{\overline{ \ A_y^x \ }^i, \overline{ \ \exists !y \ }^i}$}
\noLine
\UnaryInfC{$\Pi$}
\noLine
\UnaryInfC{$C$}
\RightLabel{$_i$}
\LeftLabel{$\exists E:$ \ }
\BinaryInfC{$C$}
\DisplayProof
\end{center} 

\noindent where in $\exists I$, $t$ is free for $x$ in $A$; and in $\exists E$, $y$ is not free in $C$ nor in any undischarged assumption of $\Pi$ except $A_y^x$ and $\exists ! y$, and either $y$ is the same as $x$ or $y$ is not free in $A$. 

The existence predicate also appears in the premise of the introduction rule for identity; the elimination rule for $=$ is Leibniz' Law: 

\begin{center}
\AxiomC{$\exists ! t$}
\LeftLabel{$= I^n:$ \ }
\UnaryInfC{$t=t$}
\DisplayProof\qquad\qquad
\AxiomC{$t_1=t_2$}
\AxiomC{$A_{t_1}^x$}
\LeftLabel{$= E$: \ } 
\BinaryInfC{$A_{t_2}^x$}
\DisplayProof
\end{center} 

\noindent where $A$ is an atomic formula and, to exclude vacuous applications of $=E$, we can require that $x$ occurs in $A$ and that $t_1$ and $t_2$ are different.

Finally, there is the rule of atomic denotation: 

\begin{prooftree}
\AxiomC{$At_1\ldots t_n$}
\LeftLabel{$AD:$ \ }
\UnaryInfC{$\exists ! t_i$}
\end{prooftree}

\noindent where $A$ is an $n$-place predicate letter (including identity) and $1\leq i\leq n$. $AD$ captures the semantic intuition that an atomic sentence can only be true if the terms that occur in it refer. 

$\mathbf{INF}^\iota$ has in addition the binary quantifier $\iota$ with the following rules: 

\begin{prooftree}
\AxiomC{$F^x_t$}
\AxiomC{$G^x_t$}
\AxiomC{$\exists !t$}
\AxiomC{$\underbrace{\overline{ \ F_z^x\ }^i, \ \overline{ \ \exists !z \ }^i}$}
\noLine
\UnaryInfC{$\Pi$}
\noLine
\UnaryInfC{$z=t$}
\RightLabel{$_i$}
\LeftLabel{$\iota I: \qquad$}
\QuaternaryInfC{$\iota x[F, G]$}
\end{prooftree}

\noindent where $t$ is free for $x$ in $F$ and in $G$, and $z$ is different from $x$, not free in $t$ and does not occur free in any undischarged assumption in $\Pi$ except $F_z^x$ and $\exists ! z$. 

\begin{prooftree}
\AxiomC{$\iota x[F, G]$}
\AxiomC{$\underbrace{\overline{ \ F_z^x\ }^i, \ \overline{ \ G_z^x \ }^i, \ \overline{ \ \exists !z \ }^i}$}
\noLine
\UnaryInfC{$\Pi$}
\noLine
\UnaryInfC{$C$}
\LeftLabel{$\iota E^1: \quad$}
\RightLabel{$_i$}
\BinaryInfC{$C$}
\end{prooftree}

\noindent where $z$ is not free in $C$ nor in any undischarged assumption of $\Pi$ except $F_z^x$, $G_z^x$ and $\exists ! z$, and either $z$ is the same as $x$ or it is not free in $F$ nor in $G$. 

\begin{prooftree}
\AxiomC{$\iota x[F, G]$}
\AxiomC{$\exists !t_1$}
\AxiomC{$\exists ! t_2$}
\AxiomC{$F_{t_1}^x$}
\AxiomC{$F_{t_2}^x$}
\LeftLabel{$\iota E^2: \quad$}
\QuinaryInfC{$t_1=t_2$}
\end{prooftree}

\noindent where $t_1$ and $t_2$ are free for $x$ in $F$. 

$\mathbf{INF}^\iota$ formalises a Russellian theory of definite descriptions, as $\iota x[F, G]$ and $\exists x (F\land\forall y(F_y^x\rightarrow x=y)\land G)$ are interderivable.

\section{Comparison of $\mathbf{INF}^\iota$ with Tennant's System}
To formalise definite descriptions using a term forming $\iota$ operator within intuitionist negative free logic, Tennant adds introduction and elimination rules for formulas of the form $\iota xF=t$ to $\mathbf{INF}$: 

\begin{prooftree} 
\AxiomC{$\exists ! t$}
\AxiomC{}
\RightLabel{$_i$}
\UnaryInfC{$z=t$}
\noLine
\UnaryInfC{$\Xi$}
\noLine
\UnaryInfC{$F_z^x$}
\AxiomC{$\underbrace{\overline{ \ F_z^x\ }^i, \ \overline{ \ \exists !z \ }^i}$}
\noLine
\UnaryInfC{$\Pi$}
\noLine
\UnaryInfC{$z=t$}
\LeftLabel{$\iota I^T:$ \ }
\RightLabel{$_i$}
\TrinaryInfC{$\iota xF=t$}
\end{prooftree} 

\noindent where in $\Xi$, $z$ does not occur in any undischarged assumption except $z=t$, and either $z$ is the same as $x$ or it is not free in $F$; and in $\Pi$, $z$ does not occur in any undischarged assumption except $F_z^x$ and $\exists ! z$.\bigskip

\begin{center} 
\AxiomC{$\iota xF=t$}
\AxiomC{$u=t$}
\LeftLabel{$\iota E^{1T}: \ $} 
\BinaryInfC{$F_u^x$}
\DisplayProof\qquad
\AxiomC{$\iota xF=t$}
\AxiomC{$F_u^x$}
\AxiomC{$\exists ! u$}
\LeftLabel{$\iota E^{2T}:$ \ }
\TrinaryInfC{$u=t$}
\DisplayProof

\bigskip

\AxiomC{$\iota xF=t$}
\LeftLabel{$\iota E^{3T}: \ $} 
\UnaryInfC{$\exists !t$}
\DisplayProof
\end{center} 

\noindent where $u$ is free for $x$ in $F$. 

It is fairly evident that there are reduction procedures for removing maximal formulas of the form $\iota xF=t$ from deductions. $\iota E^{3T}$ is a special case of the rule of atomic denotation $AD$. Notice however that it is more properly regarded as an elimination rule for $\iota$, as there is a reduction procedure for maximal formulas of the form $\iota xF=t$ that have been concluded by $\iota I^T$ and are premise of $\iota E^{3T}$. 

When negation is applied to $G(\iota xF)$, an ambiguity arises: is $\neg$ an internal negation, so that $\neg G(\iota xF)$ means `The $F$ is not $G$', or is it an external negation, so that the formula means `It is not the case that the $F$ is $G$'? Conventions or a syntactic device are needed to disambiguate. The language of Tennant's system makes no provision for distinguishing different scopes of negation. For this reason, in this section I shall restrict consideration to cases in which terms of the form $\iota xF$ occur to the left or right of $=$. I will consider a more complete system after the comparison of a restricted version of $\mathbf{INF}^\iota$ with Tennant's system. 

It might be worth noting that there is a sense in which it suffices to consider occurrences of $\iota$ terms to the left or right of identity. Whenever we are tempted to use a formula $G(\iota xF)$, we can introduce a new individual constant $c$ and use $G(c)$ and $\iota xF=c$ instead. Furthermore, in negative free logic, if $G$ is a predicate letter, then $G(\iota xF)$ can be interpreted as $\exists y(G(y)\land \iota xF=y)$, and instead of the former, we can use the latter.\footnote{In positive free logic, only half of the insinuated equivalence holds, if predicates are allowed to form sentences from $\iota$ terms: then $\exists y(G(y)\land \iota xF=y)$ implies $G(\iota xF)$, but not conversely.} There is also no need to apply the existence predicate to $\iota$ terms, as instead of $\exists !\iota xA$ we can use $\exists y\ \iota xA=y$. 

It is generally agreed that the minimal condition on a formalisation of a term forming $\iota$ operator is that it should obey \emph{Lambert's Law}:

\lbp{LL}{$LL$}{$\forall y(\iota xF=y \leftrightarrow \forall x(F\leftrightarrow x=y))$}

\noindent Tennant's rules for $\iota$ are Lambert's Law cast in the form of natural deduction. 

Call $\mathbf{INF}$ with its language modified to contain a term forming $\iota$ operator restricted to occurrences to the left or right of $=$ and augmented by Tennant's rules $\mathbf{INF}^T$. Call the same modified system augmented by Lambert's Law as an axiom $\mathbf{INF}^{LL}$.

Under the current proposal of treating $\iota$ as a binary quantifier, where `The $F$ is $G$' is formalised as $\iota x[F, G]$, formulas of the form $\iota xF=t$ employing the term forming $\iota$ operator, which intuitively mean `The $F$ is identical to $t$', can be rendered as $\iota x[F, x=t]$. Treating $\iota x[F, x=t]$ and $\iota xF=t$ as notational variants, it is not difficult to show that $\mathbf{INF}^T$ is equivalent to the fragment of $\mathbf{INF}^\iota$ where the $G$ of $\iota x[F, G]$ is restricted to identity. Call the latter system $\mathbf{INF}^{\iota R}$. For clarity, I will refer to the rules for the binary quantifier $\iota$ restricted to suit $\mathbf{INF}^{\iota R}$ by $\iota I^R$, $\iota E^{1R}$ and $\iota E^{2R}$. 

It is now convenient to have rules for the biconditional $\leftrightarrow$:\bigskip 

\begin{center}
\AxiomC{}
\RightLabel{$_i$}
\UnaryInfC{$A$}
\noLine
\UnaryInfC{$\Pi$}
\noLine
\UnaryInfC{$B$}
\AxiomC{}
\RightLabel{$_i$}
\UnaryInfC{$B$}
\noLine
\UnaryInfC{$\Pi$}
\noLine
\UnaryInfC{$A$}
\LeftLabel{$\leftrightarrow I: \ $}
\RightLabel{$_i$}
\BinaryInfC{$A\leftrightarrow B$}
\DisplayProof

\bigskip 

\AxiomC{$A\leftrightarrow B$}
\AxiomC{$A$}
\LeftLabel{$\leftrightarrow E^1: \ $}
\BinaryInfC{$B$}
\DisplayProof\quad
\AxiomC{$A\leftrightarrow B$}
\AxiomC{$B$}
\LeftLabel{$\leftrightarrow E^2: \ $}
\BinaryInfC{$A$}
\DisplayProof
\end{center} 

\noindent For perspicuity, we will mark applications of the rules for the biconditional, of Tennant's rules for $\iota$, and of $\iota I^R$, $\iota E^{1R}$ and $\iota E^{2R}$ in the deductions to follow in the next paragraphs; unmarked inferences are by the more familiar rules of $\mathbf{INF}$. 

To show that $\mathbf{INF}^T$ is a subsystem of $\mathbf{INF}^{LL}$, we observe that, treating formulas of the form $\iota xF=t$ as atomic, $\iota E^{3T}$ is a special case of $AD$, and that $\iota E^{1T}$ and $\iota E^{2T}$ are derivable from \rf{LL} by $\leftrightarrow E^1$. The following construction shows that $\iota I^T$ is also a derived rule of $\mathbf{INF}^{LL}$: 

\begin{prooftree}
\AxiomC{}
\RightLabel{$_1$}
\UnaryInfC{$z=t$}
\noLine
\UnaryInfC{$\Xi$}
\noLine
\UnaryInfC{$F_z^x$}
\AxiomC{$\underbrace{\overline{ \ F_z^x\ }^1, \ \overline{ \ \exists !z \ }^2}$}
\noLine
\UnaryInfC{$\Pi$}
\noLine
\UnaryInfC{$z=t$}
\RightLabel{$_{1 \ \leftrightarrow I} $}
\BinaryInfC{$F_z^x\leftrightarrow z=t$}
\RightLabel{$_2$}
\UnaryInfC{$\forall x(F\leftrightarrow x=t)$}
\AxiomC{$(LL)$}
\AxiomC{$\exists !t$}
\BinaryInfC{$\iota xF=t\leftrightarrow \forall x(F\leftrightarrow x=t)$}
\RightLabel{$_{\leftrightarrow E^2} $}
\BinaryInfC{$\iota xF=t$}
\end{prooftree}

\noindent Hence $\mathbf{INF}^T$ is a subsystem of $\mathbf{INF}^{LL}$. 

The next three paragraphs show that, if we write $\iota xF=t$ for $\iota x[F, x=t]$, the rules $\iota I^R$, $\iota E^{1R}$ and $\iota E^{2R}$ of $\mathbf{INF}^{\iota R}$ are derived rules of $\mathbf{INF}^T$.\bigskip  

\noindent 1. Due to the restriction on $\mathbf{INF}^{\iota R}$, applications of $\iota I^R$ are those cases of $\iota I$ in which $G_t^x$ is an identity. So it can be any identity in which $x$ is replaced by $t$ and the other term is arbitrary, i.e. any identity $(x=u)_t^x$ or $t=u$ for short: 

\begin{prooftree}
\AxiomC{$F^x_t$}
\AxiomC{$t=u$}
\AxiomC{$\exists !t$}
\AxiomC{$\underbrace{\overline{ \ F_z^x\ }^i, \ \overline{ \ \exists !z \ }^i}$}
\noLine
\UnaryInfC{$\Pi$}
\noLine
\UnaryInfC{$z=t$}
\RightLabel{$_{i \ \iota I^R}$}
\QuaternaryInfC{$\iota x[F, x=t]$}
\end{prooftree}

\noindent To derive the rule it suffices to change notation and write $\iota xF=t$ instead of $\iota x [F, x=t]$, and to observe that $F_t^x, z=t\vdash F_z^x$ by Leibniz' Law and apply $\iota I^T$: 

\begin{prooftree}
\AxiomC{$\exists !t$}
\AxiomC{$F_t^x$}
\AxiomC{}
\RightLabel{$_i$}
\UnaryInfC{$z=t$}
\BinaryInfC{$F_z^x$}
\AxiomC{$\underbrace{\overline{ \ F_z^x\ }^i, \ \overline{ \ \exists !z \ }^i}$}
\noLine
\UnaryInfC{$\Pi$}
\noLine
\UnaryInfC{$z=t$}
\RightLabel{$_{i \ \iota I^T}$}
\TrinaryInfC{$\iota xF=t$}
\end{prooftree}

\noindent The premise $t=u$ of $\iota I^R$ is redundant: a suitable identity can always be provided by deriving $(x=t)_t^x$, i.e. $t=t$, from the first premise $\exists ! t$ by $=I^n$.\bigskip 

\noindent 2. $\iota E^{1R}$ is derivable by changing notation and applying $\exists E$ with the major premise $\exists x(Fx\land x=t)$ derived from $\iota xF=t$ by $\iota E^{1T}$, multiple applications of $=I^n$ and $\iota E^{T3}$, and $\exists I$: 

\begin{prooftree} 
\AxiomC{$\iota xF=t$}
\AxiomC{$\iota xF=t$}
\UnaryInfC{$\exists ! t$}
\UnaryInfC{$t=t$}
\RightLabel{$_{\iota E^{1T}}$}
\BinaryInfC{$Ft$}
\AxiomC{$\iota xF=t$}
\UnaryInfC{$\exists ! t$}
\UnaryInfC{$t=t$}
\BinaryInfC{$Ft\land t=t$}
\AxiomC{$\iota xF=t$}
\UnaryInfC{$\exists ! t$}
\BinaryInfC{$\exists x(Fx\land x=t)$}
\end{prooftree} 

\noindent For a more elegant deduction that does not make the detour through introducing and eliminating $\exists x(Fx\land x=t)$, given a deduction $\Pi$ of $C$ from $F_x^z$, $z=t$ and $\exists !z$, replace $z$ with $t$ throughout $\Pi$, and add deductions of $\iota xF=t\vdash Ft$, $\iota xF=t\vdash t=t$ and $\iota xF=t\vdash \exists !t$ to derive the three open premises.\bigskip 

\noindent 3. Change of notation and two applications of $\iota E^{2T}$ and one of Leibniz' Law derive $\iota E^{2R}$: 

\begin{prooftree}
\AxiomC{$\iota xF=t$}
\AxiomC{$\exists ! t_1$}
\AxiomC{$F_{t_1}^x$}
\RightLabel{$_{\iota E^{2T}}$}
\TrinaryInfC{$t=t_1$}
\AxiomC{$\iota xF=t$}
\AxiomC{$\exists ! t_2$}
\AxiomC{$F_{t_2}^x$}
\RightLabel{$_{\iota E^{2T}}$}
\TrinaryInfC{$t=t_2$}
\BinaryInfC{$t_1=t_2$}
\end{prooftree}

\noindent Thus $\mathbf{INF}^{\iota R}$ is a subsystem of $\mathbf{INF}^T$.

\begin{landscape} 
Finally, we derive \rf{LL} in the version appropriate to $\mathbf{INF}^{\iota R}$, i.e. with $\iota xA=y$ replaced by $\iota x[A, x=y]$: 

\lbp{LL'}{$LL'$}{Lambert's Law: $\forall y(\iota x[A, x=y] \leftrightarrow \forall x(A\leftrightarrow x=y))$}

\noindent 1. $\iota x[A, x=y]\vdash \forall x(A\leftrightarrow x=y)$

\begin{prooftree}
\AxiomC{$\iota x[A, x=y]$}
\AxiomC{$\iota x[A, x=y]$}
\AxiomC{}
\RightLabel{$_3$}
\UnaryInfC{$\exists ! x$}
\AxiomC{}
\RightLabel{$_4$}
\UnaryInfC{$\exists ! z$}
\AxiomC{}
\RightLabel{$_4$}
\UnaryInfC{$z=y$}
\BinaryInfC{$\exists ! y$}
\AxiomC{}
\RightLabel{$_4$}
\UnaryInfC{$A_z^x$}
\AxiomC{}
\RightLabel{$_4$}
\UnaryInfC{$z=y$}
\BinaryInfC{$A_y^x$}
\AxiomC{}
\RightLabel{$_2$}
\UnaryInfC{$A$}
\RightLabel{$_{\iota E^{2 R}} $}
\QuinaryInfC{$x=y$}
\AxiomC{$\iota x[A, x=y]$}
\AxiomC{}
\RightLabel{$_1$}
\UnaryInfC{$A_z^x$}
\AxiomC{}
\RightLabel{$_1$}
\UnaryInfC{$z=y$}
\BinaryInfC{$A_y^x$}
\AxiomC{}
\RightLabel{$_2$}
\UnaryInfC{$x=y$}
\BinaryInfC{$A$}
\RightLabel{$_{1 \ \iota E^{1 R}} $}
\BinaryInfC{$A$}
\RightLabel{$_{2 \ \leftrightarrow I}$}
\BinaryInfC{$A \leftrightarrow x=y$}
\RightLabel{$_3$}
\UnaryInfC{$\forall x(A\leftrightarrow x=y)$}
\RightLabel{$_{4 \ \iota E^{1R}}$}
\BinaryInfC{$\forall x(A\leftrightarrow x=y)$}
\end{prooftree} 

\bigskip

\noindent 2. $\forall x(A\leftrightarrow x=y), \exists ! y\vdash\iota x[A, x=y]$ 

\begin{prooftree} 
\AxiomC{$\forall x(A\leftrightarrow x=y)$} 
\AxiomC{$\exists ! y$}
\BinaryInfC{$A_y^x\leftrightarrow y=y$}
\AxiomC{$\exists ! y$}
\UnaryInfC{$y=y$}
\BinaryInfC{$A_y^x$}
\AxiomC{$\exists ! y$}
\UnaryInfC{$y=y$}
\AxiomC{$\exists ! y$}
\AxiomC{$\forall x(A\leftrightarrow x=y)$} 
\AxiomC{}
\RightLabel{$_1$}
\UnaryInfC{$\exists ! z$}
\BinaryInfC{$A_z^x\leftrightarrow z=y$}
\AxiomC{}
\RightLabel{$_1$}
\UnaryInfC{$A_z^x$}
\BinaryInfC{$z=y$}
\RightLabel{$_{1 \ \iota I^R}$}
\QuaternaryInfC{$\iota x[A, x=y]$}
\end{prooftree} 

\noindent Now from 1 and 2 by $\leftrightarrow I$, we have $\exists ! y\vdash \iota x[A, x=y]\leftrightarrow \forall x(A\leftrightarrow x=y)$, and so by $\forall I$, $\vdash \forall y(\iota x[A, x=y] \leftrightarrow \forall x(A\leftrightarrow x=y))$.\bigskip

Hence $\mathbf{INF}^{LL}$ is a subsystem of $\mathbf{INF}^{\iota R}$. This completes the circle, and we have shown:

\begin{theorem}
$\mathbf{INF}^T$, $\mathbf{INF}^{LL}$ and $\mathbf{INF}^{\iota R}$ are equivalent. 
\end{theorem}
\end{landscape}

\section{Comparison of $\mathbf{INF}^\iota$ with an Intuitionist Version of a System of Lambert's}
As noted towards the beginning of the previous section, in the absence of a formal device or a convention for distinguishing two ways of applying negation to $G(\iota xF)$, $\neg G(\iota xF)$ is ambiguous: $\neg$ can either be internal or external negation. To eliminate ambiguity, Lambert introduces an abstraction operator $\Delta$ that forms complex predicate terms $\Delta xB$ from open formulas $B$, and with the formation rule that if $\Delta x B$ is a predicate term and $t$ an individual term, then $\Delta xB, t$ is a formula. Semantically, $\Delta xB, t$ is interpreted as true just in case $t$ exists and $Bt$ is true.\footnote{For this and the following, see \citep[39ff]{lambertfreedef}.} In this section I will compare $\mathbf{INF}^\iota$ to an intuitionist version of Lambert's system. Like Lambert, I will only consider unary predicates and keep the discussion fairly informal.\footnote{Lambert provides a more general treatment of an abstraction operator in classical positive free logic, but without a description operator, in \citep{lambertposfreelogpreabstr}. A more complete and precise comparison of my treatment of definite description with Lambert's is reserved for sequels to this paper on the binary quantifier $\iota$ in intuitionist positive free logic and in negative and positive classical free logic. Fitting and Mendelsohn also employ predicate abstraction as a device for distinguishing scope within modal logic \citep[Ch 12]{mendelsohnfitting}.} 

In Lambert's system, $\Delta$ is governed by a principle regarded either as an axiom or as a contextual definition: 

\lbp{Ab}{$\Delta t$}{$\Delta xB, t \leftrightarrow (\exists !t \land B_t^x)$ \qquad ($t$ free for $x$ in $B$ and $x$ not free in $t$)} 

\noindent To formalise a free Russellian theory of definite descriptions, Lambert adds Lambert's Law and the following principle to negative free logic, also regarded either as an axiom or as a contextual definition: 

\lbp{Ru}{$\Delta\iota$}{$\Delta xB, \iota xA \leftrightarrow \exists z(\iota xA=z\land B_z^x)$} 

\noindent Lambert uses a classical negative free logic, but in this section I will consider adding \rf{LL}, \rf{Ab} and \rf{Ru} to $\mathbf{INF}$. Call the resulting system $\mathbf{INF}^{LL\Delta}$. In this system, what we may call the \emph{primary occurrences} of $\iota$ terms are those to the left or right of identity and which are governed by Lambert's Law. What we may call the \emph{secondary occurrences} of $\iota$ terms are those introduced on the basis of the primary ones by the contextual definition \rf{Ru}.

Lambert notes three characteristically Russellian theorems that are consequences of \rf{LL}, \rf{Ab} and \rf{Ru}: 

\lbp{R1}{$R1$}{$\exists ! \iota xA\leftrightarrow \exists y\forall x (A\leftrightarrow x=y)$}

\lbp{R2}{$R2$}{$\Delta xB, \iota yA \leftrightarrow \exists z(\forall y (A\leftrightarrow y=z)\land B_z^x)$}

\lbp{R3}{$R3$}{$\iota xA=t \rightarrow A_t^x$ \qquad ($t$ free for $x$ in $A$ and $x$ not free in $t$)}

\noindent A further characteristically Russellian thesis mentioned by Morscher and Simons \citep[19]{morschersimonsfreelogic} is worth listing: 

\lbp{R4}{$R4$}{$\exists ! \iota x A \rightarrow A(\iota xA)$}

\noindent We will show that $\mathbf{INF}^{LL\Delta}$ and $\mathbf{INF}^\iota$ are equivalent, and then, to take a convenient opportunity to illustrate the workings of the latter system, derive formulas corresponding to \rf{R1} to \rf{R4} in $\mathbf{INF}^\iota$. 

In the present formalisation of $\iota$ as a binary quantifier, no conventions or syntactic devices are needed for the disambiguation of complex formulas involving $\iota$. Ambiguity is avoided by the notation for the operator itself, which incorporates the relevant scope distinction. In this sense, the current formalisation of definite descriptions is more versatile than a formalisation using a term forming operator: it does the work of both, the term forming $\iota$ operator and the abstraction operator. 

There is a certain redundancy in Lambert's axioms. $\exists !t\land B_t^x$ is equivalent to $\exists z (t=z\land B_z^x)$:\footnote{The second deduction is constructed so as not to appeal to any rules of $\mathbf{INF}$ that are not also rules of the system $\mathbf{IPF}$ of \citep[Sec 3]{kurbisiotaI}. The first deduction can be adjusted to $\mathbf{IPF}$ by deducing $t=t$ from no premises by $=I$.}

\begin{prooftree}
\AxiomC{$\exists ! t\land B_t^x$}
\UnaryInfC{$\exists !t$}
\UnaryInfC{$t=t$}
\AxiomC{$\exists ! t\land B_t^x$}
\UnaryInfC{$B_t^x$}
\BinaryInfC{$t=t\land B_t^x$}
\AxiomC{$\exists ! t\land B_t^x$}
\UnaryInfC{$\exists !t$}
\BinaryInfC{$\exists z(t=z\land B_z^x)$}
\end{prooftree}

\begin{prooftree}
\AxiomC{$\exists z(t=z\land B_z^x)$}
\AxiomC{}
\RightLabel{$_1$}
\UnaryInfC{$t=z\land B_z^x$}
\UnaryInfC{$t=z$}
\AxiomC{}
\RightLabel{$_1$}
\UnaryInfC{$\exists ! z$}
\BinaryInfC{$\exists !t$}
\AxiomC{}
\RightLabel{$_1$}
\UnaryInfC{$t=z\land B_z^x$}
\UnaryInfC{$t=z$}
\AxiomC{}
\RightLabel{$_1$}
\UnaryInfC{$t=z\land B_z^x$}
\UnaryInfC{$B_z^x$}
\BinaryInfC{$B_t^x$}
\BinaryInfC{$\exists ! t\land B_t^x$}
\RightLabel{$_1$}
\BinaryInfC{$\exists ! t\land B_t^x$}
\end{prooftree}  

\noindent This means that there is a uniform treatment of the $\Delta$ operator, irrespective of whether the term a predicate abstract is applied to is an $\iota$ term or not, and one axiom suffices to replace \rf{Ab} and \rf{Ru}: 

\lbp{Ab'}{$\Delta t'$}{$\Delta xB, t \leftrightarrow \exists z(t=z \land B_z^x)$ \qquad ($t$ free for $z$ in $B$ and $z$ not free in $t$)} 

\noindent This works only for a Russellian theory of definite descriptions, however: an alternative theory of definite descriptions within positive free logic may be intended to provide room for the option that $\Delta xB, \iota xA$ is true even though there is no unique $A$: such a theory may contain \rf{Ab} but not \rf{Ru}. 

Furthermore, $\Delta xB, t$ is equivalent to $\Delta xB, \iota x(x=t)$, both being equivalent to $\exists z(t=z\land B_z^x)$. Thus there is a sense in which nothing is lost from Lambert's system if the formation rules for the abstraction operator were reformulated so as to require a predicate and an $\iota$ term to form a formula out of them. The $\iota$ symbol, being embedded within the $\Delta$ operator, could then just as well be omitted, so that $\Delta$ forms a formula out of two predicates, which is exactly how the $\iota$ operator works in $\mathbf{INF}^\iota$. Of course what is crucial for Lambert's system is Lambert's Law, and in his formulation of it $\Delta$ does not occur. The present system is thus in a sense more economical than Lambert's. 

We can emulate Lambert's use of both, the abstraction operator and the term forming $\iota$ operator, in the present system: $\Delta xG, \iota xF$ is translated as $\iota x[F, G]$, and where $t$ is not an $\iota$ term, $\Delta xA, t$ is translated as $\iota x[t=x, A]$: instead of naming an object and applying a predicate to it, we pick out the object by a predicate that is true at most of it. Then what is expressed by $\iota xA=y$ in Lambert's system is expressed in $\mathbf{INF}^\iota$ by $\iota x[A, x=y]$, and what is expressed by $\exists !\iota xA$ is expressed by $\iota x[A, \exists ! x]$. 

A little more precisely, to show that $\mathbf{INF}^{LL\Delta}$ and $\mathbf{INF}^\iota$ are equivalent, observe that their languages differ only in that the former has $\Delta$ and the term forming $\iota$, which the latter lacks, and in that the latter has the binary quantifier $\iota$, which the former lacks. We construct a translation $\tau$ from the language of $\mathbf{INF}^{LL\Delta}$ to the language of $\mathbf{INF}^\iota$. Atomic sentences and those containing operators other than $\Delta$ and $\iota$ are translated homophonically:

\setlist{noitemsep}
\begin{itemize}[align=parleft]
\item[(a)] if $A$ is atomic formula not containing any $\iota$ terms, then $\tau(A)=A$, 

\item[(b)] if the main operator of $A$ is a unary operator $\ast$ (i.e. $\ast$ is $\neg$, $\exists$ or $\forall$), then $\tau(\ast B)=\ast \tau(B)$, 

\item[(c)] if $\ast$ is a binary sentential operator, then $\tau (A\ast B)=\tau(A)\ast\tau(B)$.
\end{itemize} 

\noindent Next, the primary occurrences of $\iota$ terms: 

\begin{itemize}[align=parleft]
\item[(d.i)] $\tau(\iota xA=t)=\iota x[\tau(A), x=t]$; similarly for $t=\iota xA$ (i.e. $\tau(\iota xA=\iota yB)=\iota x[\tau(A), \iota y[\tau(B), x=y]]$). 
\end{itemize}

\noindent For formulas containing $\Delta$ and the secondary occurrences of $\iota$ terms, we need a distinction: 

\begin{itemize}[align=parleft]
\item[(e.i)] if $t$ is not an $\iota$ term, then $\tau(\Delta xB, t)=\iota x[t=x, \tau(B)]$,

\item[(e.ii)] if $t$ is an $\iota$ term $\iota xA$, then $\tau(\Delta xB, t)=\iota x[\tau(A), \tau(B)]$.
\end{itemize}

\noindent To construct a translation $\upsilon$ from the language of $\mathbf{INF}^\iota$ to the language of $\mathbf{INF}^{LL\Delta}$, we recycle clauses (a) to (c) of $\tau$ and add only $\upsilon(\iota x[A, B])=\Delta x\upsilon(B), \iota x\upsilon(A)$, letting the contextual definitions \rf{Ab} and \rf{Ru} do the rest. 

Let $\tau(\Gamma)$, $\upsilon(\Gamma)$ be the set of formulas in $\Gamma$ translated by $\tau$, $\upsilon$. We have:

\begin{theorem}
$\mathbf{INF}^\iota$ is equivalent to $\mathbf{INF}^{LL\Delta}$: (a) if $\Gamma\vdash A$ in $\mathbf{INF}^\iota$, then $\upsilon(\Gamma)\vdash \upsilon(A)$ in $\mathbf{INF}^{LL\Delta}$; (b) if $\Gamma\vdash A$ in $\mathbf{INF}^{LL\Delta}$, then $\tau(\Gamma)\vdash \tau (A)$ in $\mathbf{INF}^\iota$.
\end{theorem}

\noindent \emph{Proof.} (a) It suffices to observe that the introduction and elimination rules for $\iota$ of $\mathbf{INF}^\iota$ remain valid under the translation $\upsilon$, due to the equivalence of $\iota x[F, G]$ with $\exists x(F\land \forall y(F_y^x\rightarrow y=x)\land G)$ and \rf{R2}. (b) It suffices to prove the translations of \rf{LL}, \rf{Ab} and \rf{Ru} under $\tau$ in $\mathbf{INF}^\iota$: 

\lbp{LLtau}{$LL^\tau$}{$\forall y(\iota x[\tau(A), x=y] \leftrightarrow \forall x(\tau(A)\leftrightarrow x=y))$} 

\lbp{Abtau}{$\Delta t^\tau$}{$\iota x[x=t, \tau(A)]\leftrightarrow ( \exists ! t\land \tau(A)_t^x)$ \qquad ($t$ free for $x$ in $\tau(A)$ and $x$ not free in $t$)} 

\lbp{Rutau}{$\Delta \iota^\tau$}{$\iota x[\tau(A), \tau(B)]\leftrightarrow \exists z(\iota x[\tau(A), x=z] \land \tau(B)_z^x)$}

\noindent For readability I will prove these equivalences `schematically', it being understood that the formulas $A$ and $B$ in the deductions to follow are translations under $\tau$.\footnote{From an alternative perspective, the provability of these equivalences shows that adding \rf{LL}, \rf{Ab} and \rf{Ru} to $\mathbf{INF}^\iota$ does not increase its expressive power, as for each formula containing the term forming $\iota$ operator and $\Delta$, there is a provably equivalent one containing only the binary quantifier $\iota$.} Then \rf{LLtau} is \rf{LL'}, which we proved earlier. The other two we prove next.\bigskip 

\noindent \rf{Abtau} \ $\iota x[x=t, A]\leftrightarrow ( \exists ! t\land A_t^x)$ \qquad ($t$ free for $x$ in $A$ and $x$ not free in $t$)\bigskip

\noindent 1. $\iota x[x=t, A]\vdash \exists ! t\land A_t^x$ 

\begin{prooftree} 
\AxiomC{$\iota x[t=x, A]$}
\AxiomC{}
\RightLabel{$_1$}
\UnaryInfC{$\exists ! x$}
\AxiomC{}
\RightLabel{$_1$}
\UnaryInfC{$t=x$}
\BinaryInfC{$\exists ! t$}
\AxiomC{}
\RightLabel{$_1$}
\UnaryInfC{$A$}
\AxiomC{}
\RightLabel{$_1$}
\UnaryInfC{$t=x$}
\BinaryInfC{$A_t^x$}
\BinaryInfC{$\exists !t\land A_t^x$}
\RightLabel{$_{1 \ \iota E^1}$}
\BinaryInfC{$\exists !t\land A_t^x$}
\end{prooftree} 

\noindent 2. $\exists ! t\land A_t^x\vdash \iota x[x=t, A]$\bigskip

\begin{prooftree}
\AxiomC{$\exists ! t\land A_t^x$}
\UnaryInfC{$\exists !t$}
\UnaryInfC{$t=t$}
\AxiomC{$\exists ! t\land A_t^x$}
\UnaryInfC{$A_t^x$}
\AxiomC{$\exists ! t\land A_t^x$}
\UnaryInfC{$\exists !t$}
\AxiomC{}
\RightLabel{$_1$}
\UnaryInfC{$z=t$}
\RightLabel{$_{1 \ \iota I}$}
\QuaternaryInfC{$\iota x[x=t, A]$}
\end{prooftree} 

\noindent This is a correct application of $\iota I$: $F_t^x$ is $(x=t)_t^x$, i.e. $t=t$, and $F_z^x$ is $(x=t)_z^x$, i.e. $z=t$. $\exists ! z$ is discharged vacuously.\bigskip

\begin{landscape}
\noindent \rf{Rutau} \ $\iota x[A, B]\leftrightarrow \exists z(\iota x[A, x=z] \land B_z^x)$\bigskip

\noindent 1. $\iota x[A, B]\vdash\exists z(\iota x[A, x=z]\land B_z^x)$ 

\begin{prooftree}
\AxiomC{$\iota x[A, B]$}
\AxiomC{}
\RightLabel{$_2$}
\UnaryInfC{$A_z^x$}
\AxiomC{}
\RightLabel{$_2$}
\UnaryInfC{$\exists ! z$}
\UnaryInfC{$z=z$}
\AxiomC{}
\RightLabel{$_2$}
\UnaryInfC{$\exists ! z$}
\AxiomC{$\iota x[A, B]$}
\AxiomC{}
\RightLabel{$_1$}
\UnaryInfC{$\exists !x$}
\AxiomC{}
\RightLabel{$_2$}
\UnaryInfC{$\exists !z$}
\AxiomC{}
\RightLabel{$_1$}
\UnaryInfC{$A$}
\AxiomC{}
\RightLabel{$_2$}
\UnaryInfC{$A_z^x$}
\RightLabel{$_{\iota E^2}$}
\QuinaryInfC{$x=z$}
\RightLabel{$_{1 \ \iota I}$}
\QuaternaryInfC{$\iota x[A, x=z]$}
\AxiomC{}
\RightLabel{$_2$}
\UnaryInfC{$B_z^x$}
\BinaryInfC{$\iota x[A, x=z]\land B_z^x$}
\AxiomC{}
\RightLabel{$_2$}
\UnaryInfC{$\exists !z$}
\BinaryInfC{$\exists z(\iota x[A, x=z]\land B_z^x)$}
\RightLabel{$_{2 \ \iota E^1}$}
\BinaryInfC{$\exists z(\iota x[A, x=z]\land B_z^x)$}
\end{prooftree} 

\noindent 2. $\exists z(\iota x[A, x=z]\land B_z^x)\vdash \iota x[A, B]$\bigskip

\noindent First, $\iota x[A, x=z], B_z^x \vdash \iota x[A, B]$: 

\begin{prooftree}
\AxiomC{$\iota x[A, x=z]$} 
\AxiomC{}
\RightLabel{$_2$}
\UnaryInfC{$A$}
\AxiomC{}
\RightLabel{$_2$}
\UnaryInfC{$x=z$}
\BinaryInfC{$A_z^x$}
\AxiomC{$B_z^x$}
\AxiomC{}
\RightLabel{$_2$}
\UnaryInfC{$\exists ! z$}
\AxiomC{$\iota x[A, x=z]$}
\AxiomC{}
\RightLabel{$_1$}
\UnaryInfC{$\exists !y$}
\AxiomC{}
\RightLabel{$_2$}
\UnaryInfC{$\exists !z$}
\AxiomC{}
\RightLabel{$_1$}
\UnaryInfC{$A_y^x$}
\AxiomC{}
\RightLabel{$_2$}
\UnaryInfC{$A$}
\AxiomC{}
\RightLabel{$_2$}
\UnaryInfC{$x=z$}
\BinaryInfC{$A_z^x$}
\RightLabel{$_{\iota E^2}$}
\QuinaryInfC{$y=z$}
\RightLabel{$_{1 \ \iota I}$}
\QuaternaryInfC{$\iota x[A, B]$} 
\RightLabel{$_{2 \ \iota E^1}$}
\BinaryInfC{$\iota x[A, B]$} 
\end{prooftree}

\noindent Thus $\iota x[A, x=z]\land B_z^x\vdash \iota x[A, B]$, and so $\exists z(\iota x[A, x=z]\land B_z^x)\vdash \iota x[A, B]$. In this last application of $\exists E$, $\exists !z$ is discharged vacuously. Notice that it would have been possible to discharge only one (or indeed none) of the $\exists !z$ by $\iota E^2$, and the discharge the other (or both) by the application of $\exists E$.\bigskip

\noindent This completes the proof of Theorem 2. 
\end{landscape} 

Under translation $\tau$, \rf{R1}, \rf{R2}, \rf{R3} and \rf{R4} become:

\lbp{R1'}{$R1^\tau$}{$\iota x[\tau(A), \exists ! x]\leftrightarrow\exists y\forall x (\tau(A)\leftrightarrow x=y)$} 

\lbp{R2'}{$R2^\tau$}{$\iota x[\tau(A), \tau(B)]\leftrightarrow \exists z(\forall y (\tau(A)\leftrightarrow y=z)\land \tau(B)_z^x)$}

\lbp{R3'}{$R3^\tau$}{$\iota x[\tau(A), x=t]\rightarrow \tau(A)_t^x$ \qquad ($t$ free for $x$ in $\tau(A)$ and $x$ not free in $t$)}

\lbp{R4'}{$R4^\tau$}{$\iota x[\tau(A), \exists ! x]\rightarrow \iota x[\tau(A), \tau(A)]$}

\noindent  \rf{R2'} follows from the interderivability of  $\exists x(A\land \forall y(A_y^x\rightarrow x=y)\land B)$ with $\iota x[A, B]$ (see \citep[90f]{kurbisiotaI}). The rest are proved on the following pages, once more `schematically' and with $\tau$ suppressed for readability. The proofs presuppose a judicious choice of variables. 

\begin{landscape} 
\noindent \rf{R1'} \ $\iota x[A, \exists ! x]\leftrightarrow\exists y\forall x (A\leftrightarrow x=y)$\bigskip

\noindent 1. $\iota x[A, \exists ! x]\vdash\exists y\forall x (A\leftrightarrow x=y)$

\begin{prooftree}
\AxiomC{$\iota x[A, \exists ! x]$}
\AxiomC{$\iota x[A, \exists ! x]$}
\AxiomC{}
\RightLabel{$_2$}
\UnaryInfC{$\exists ! x$}
\AxiomC{}
\RightLabel{$_3$}
\UnaryInfC{$\exists ! y $}
\AxiomC{}
\RightLabel{$_1$}
\UnaryInfC{$A$}
\AxiomC{}
\RightLabel{$_3$}
\UnaryInfC{$A_y^x$}
\RightLabel{$_{\iota E^2}$}
\QuinaryInfC{$x=y$}
\AxiomC{}
\RightLabel{$_3$}
\UnaryInfC{$A_y^x$}
\AxiomC{}
\RightLabel{$_1$}
\UnaryInfC{$x=y$}
\BinaryInfC{$A$}
\RightLabel{$_{1 \ \leftrightarrow I}$}
\BinaryInfC{$A\leftrightarrow x=y$}
\RightLabel{$_2$}
\UnaryInfC{$\forall x(A\leftrightarrow x=y)$}
\AxiomC{}
\RightLabel{$_3$}
\UnaryInfC{$\exists ! y$}
\BinaryInfC{$\exists y\forall x(A\leftrightarrow x=y)$}
\RightLabel{$_{3 \ \iota E^1}$}
\BinaryInfC{$\exists y\forall x(A\leftrightarrow x=y)$}
\end{prooftree}

\noindent 2. $\exists y\forall x (A\leftrightarrow x=y)\vdash\iota x[A, \exists ! x]$

\begin{prooftree}
\AxiomC{$\exists y\forall x (A\leftrightarrow x=y)$}
\AxiomC{}
\RightLabel{$_2$}
\UnaryInfC{$\forall x (A\leftrightarrow x=y)$}
\AxiomC{}
\RightLabel{$_2$}
\UnaryInfC{$\exists ! y$}
\BinaryInfC{$A_y^x\leftrightarrow y=y$}
\AxiomC{}
\RightLabel{$_2$}
\UnaryInfC{$\exists ! y$}
\UnaryInfC{$y=y$}
\RightLabel{$_{\leftrightarrow E^2}$}
\BinaryInfC{$A_y^x$}
\AxiomC{}
\RightLabel{$_2$}
\UnaryInfC{$\exists !_y^x$} 
\AxiomC{}
\RightLabel{$_2$}
\UnaryInfC{$\exists ! y$}
\AxiomC{}
\RightLabel{$_2$}
\UnaryInfC{$\forall x (A\leftrightarrow x=y)$}
\AxiomC{}
\RightLabel{$_1$}
\UnaryInfC{$\exists ! v$}
\BinaryInfC{$A_v^x\leftrightarrow v=y$}
\AxiomC{}
\RightLabel{$_1$}
\UnaryInfC{$A_v^x$}
\RightLabel{$_{\leftrightarrow E^1}$}
\BinaryInfC{$v=y$}
\RightLabel{$_{1 \ \iota I}$}
\QuaternaryInfC{$\iota x[A, \exists ! x]$}
\RightLabel{$_2$}
\BinaryInfC{$\iota x[A, \exists !x]$}
\end{prooftree}
\end{landscape}

\begin{landscape} 
\noindent \rf{R3'} \ $\iota x[A, x=t]\rightarrow A_t^x$ \qquad ($t$ free for $x$ in $A$ and $x$ not free in $t$) 

\begin{prooftree}
\AxiomC{$\iota x[A, x=t]$} 
\AxiomC{}
\RightLabel{$_1$}
\UnaryInfC{$A$}
\AxiomC{}
\RightLabel{$_1$}
\UnaryInfC{$x=t$}
\BinaryInfC{$A_t^x$}
\RightLabel{$_{1 \ \iota E^1}$}
\BinaryInfC{$A_t^x$}
\end{prooftree} 

\noindent We also have $\iota x[A, x=t]\rightarrow \exists ! t$  ($x$ not free in $t$): 

\begin{prooftree}
\AxiomC{$\iota x[A, x=t]$}  
\AxiomC{}
\RightLabel{$_1$}
\UnaryInfC{$\exists ! x$}
\AxiomC{}
\RightLabel{$_1$}
\UnaryInfC{$x=t$}
\BinaryInfC{$\exists ! t$}
\RightLabel{$_{1 \ \iota E^1}$}
\BinaryInfC{$\exists !t$}
\end{prooftree} 

\noindent Hence $\iota x[A, x=t]\rightarrow (\exists ! t\land A_t^x)$, and so by \rf{Abtau}, $\iota x[A, x=t]\rightarrow \iota x[x=t, A]$. We do not, however, have the converse. $(\exists ! t\land A_t^x)\rightarrow\iota x[A, x=t]$ is not true. $\iota x[A, x=t]$ means `The $A$ is identical to $t$', and this does not follow from the existence of a $t$ which is $A$, i.e. $\exists ! t\land A_t^x$.\bigskip 

\noindent \rf{R4'} \ $\iota x[A, \exists !x]\vdash \iota x[A, A]$

\begin{prooftree}
\AxiomC{$\iota x[A, \exists !x]$}
\AxiomC{}
\RightLabel{$_2$}
\UnaryInfC{$A_z^x$}
\AxiomC{}
\RightLabel{$_2$}
\UnaryInfC{$A_z^x$}
\AxiomC{}
\RightLabel{$_2$}
\UnaryInfC{$\exists !z$}
\AxiomC{$\iota x[A, \exists ! x]$}
\AxiomC{}
\RightLabel{$_1$}
\UnaryInfC{$\exists !v$}
\AxiomC{}
\RightLabel{$_2$}
\UnaryInfC{$\exists !z$}
\AxiomC{}
\RightLabel{$_1$}
\UnaryInfC{$A_v^x$}
\AxiomC{}
\RightLabel{$_2$}
\UnaryInfC{$A_z^x$}
\RightLabel{$_{\iota E^2}$}
\QuinaryInfC{$v=z$}
\RightLabel{$_{1 \ \iota I}$}
\QuaternaryInfC{$\iota x[A, A]$}
\RightLabel{$_{2 \ \iota E^1}$}
\BinaryInfC{$\iota x[A, A]$} 
\end{prooftree}

\end{landscape}

\begin{landscape}
To close this section, a few words about $\forall E$, $\exists I$ and $=E$. In systems where $\iota$ is a term forming operator, $\iota$ terms can be used as terms instantiating universal generalisations, as terms over which to generalise existentially, and as terms to the left or right of identity in Leibniz's Law. To establish that the current system is as versatile as a system in which this is possible, it remains to be shown that these uses of $\iota$ terms can be reconstructed in the present formalism. In other words, we need to show:

\lbp{alliota}{$\forall\iota$}{$\forall xB, \iota x[A, \exists ! x]\vdash \iota x[A, B]$} 

\lbp{existsiota}{$\exists\iota$}{$\iota x[A, B], \iota x[A, \exists ! x]\vdash \exists xB$}

\lbp{=iota}{$=\iota$}{$B_t^x, \iota x[A, x=t] \vdash \iota x[A, B]$} 

\noindent An inference concluding the existence of an $\iota$ term by $AD$ is a special case of $\iota x[F, G]\vdash \iota x[F, \exists ! x]$, which holds by \rf{R2'}, \rf{R1'} and general logic. I will only show that \rf{alliota} and \rf{=iota} hold, the proof of \rf{existsiota} being similar.\bigskip

\noindent \rf{alliota} \ $\forall xB, \iota x[A, \exists ! x]\vdash \iota x[A, B]$

\begin{prooftree}
\AxiomC{$\iota x[A, \exists ! x]$} 
\AxiomC{}
\RightLabel{$_2$}
\UnaryInfC{$A_z^x$}
\AxiomC{$\forall xB$}
\AxiomC{}
\RightLabel{$_2$}
\UnaryInfC{$\exists ! z$}
\BinaryInfC{$B_z^x$}
\AxiomC{}
\RightLabel{$_2$}
\UnaryInfC{$\exists ! z$}
\AxiomC{$\iota x[A, \exists ! x]$} 
\AxiomC{}
\RightLabel{$_1$}
\UnaryInfC{$\exists ! y$}
\AxiomC{}
\RightLabel{$_2$}
\UnaryInfC{$\exists ! z$}
\AxiomC{}
\RightLabel{$_1$}
\UnaryInfC{$A_y^x$}
\AxiomC{}
\RightLabel{$_2$}
\UnaryInfC{$A_z^x$}
\RightLabel{$_{\iota E^2}$}
\QuinaryInfC{$y=z$}
\RightLabel{$_{1 \ \iota I}$}
\QuaternaryInfC{$\iota x[A, B]$}
\RightLabel{$_{2 \ \iota E^1}$}
\BinaryInfC{$\iota x[A, B]$}
\end{prooftree} 

\noindent\rf{=iota} \ $B_t^x, \iota x[A, x=t] \vdash \iota x[A, B]$ 

\begin{prooftree}
\AxiomC{$\iota x[A, x=t]$}
\AxiomC{}
\RightLabel{$_2$}
\UnaryInfC{$x=t$}
\AxiomC{}
\RightLabel{$_2$}
\UnaryInfC{$A$}
\BinaryInfC{$A_t^x$}
\AxiomC{$B_t^x$}
\AxiomC{}
\RightLabel{$_2$}
\UnaryInfC{$\exists ! x$}
\AxiomC{}
\RightLabel{$_2$}
\UnaryInfC{$x=t$}
\BinaryInfC{$\exists ! t$} 
\AxiomC{$\iota x[A, x=t]$}
\AxiomC{}
\RightLabel{$_1$}
\UnaryInfC{$\exists ! v$}
\AxiomC{}
\RightLabel{$_2$}
\UnaryInfC{$x=t$}
\AxiomC{}
\RightLabel{$_2$}
\UnaryInfC{$\exists ! x$}
\BinaryInfC{$\exists ! t$}
\AxiomC{}
\RightLabel{$_1$}
\UnaryInfC{$A_v^x$}
\AxiomC{}
\RightLabel{$_2$}
\UnaryInfC{$x=t$}
\AxiomC{}
\RightLabel{$_2$}
\UnaryInfC{$A$}
\BinaryInfC{$A_t^x$}
\RightLabel{$_{\iota E^2}$} 
\QuinaryInfC{$v=t$}
\RightLabel{$_{1 \ \iota I}$} 
\QuaternaryInfC{$\iota x[A, B]$}
\RightLabel{$_{2 \ \iota E^1}$}
\BinaryInfC{$\iota x[A, B]$} 
\end{prooftree} 
\end{landscape}

\section{Conclusion and Further Work}
The present formalism has certain advantages over the use of $\iota$ as a term forming operator. It incorporates scope distinctions within the notation, without the need for an abstraction operator or other syntactic devices or conventions. It provides a natural formalisation of a theory of definite descriptions, here developed within intuitionist negative free logic. The resulting system has desirable proof-theoretic properties, as deductions in it normalise, and it is equivalent to well known axiomatic theories of definite descriptions. 

Scope distinctions are of particular interest to the development of a theory of definite descriptions within modal logic. Fitting and Mendelsohn, for instance, provide a detailed account of definite descriptions within quantified modal logic \citep[Ch 12]{mendelsohnfitting}, which uses an abstraction operator for scope distinction. They observe that scope distinctions are already needed for formulas containing individual constants, if they are not interpreted rigidly, and so they introduce predicate abstraction well before definite descriptions. However, in their system, as in Lambert's, predicate abstraction does not appear to play any further role than marking scope distinctions. The present notation provides a perspicuous way of distinguishing the scope of modal operators that is independent of abstraction operators:\bigskip

\noindent It is possible that the $F$ is $G$: $\Diamond \iota x[F, G]$

\noindent The $F$ is possibly $G$: $\iota x[F, \Diamond G]$.

\noindent The possible $F$ is G: $\iota x[\Diamond F, G]$\bigskip

\noindent For scope distinctions with regard to non-rigidly interpreted individual constants, we can use the technique of simulating the use of a constant $t$ by a predicate $x=t$ introduced earlier. It would be worth comparing the approach proposed here with Fitting's and Mendelsohn's, but this must wait for another occasion.\bigskip

\setlength{\bibsep}{0pt}
\bibliographystyle{chicago}
\bibliography{iota}

\end{document}